\documentclass[A4paper]{article}
\usepackage{amsmath}
\usepackage{amssymb}
\usepackage{graphicx,psfrag,epsf}
\usepackage{enumerate}
\usepackage{natbib}
\usepackage{url} % not crucial - just used below for the URL 
\usepackage{fullpage}
\usepackage{MnSymbol}
\usepackage{bm}
\usepackage{tikz}
\usepackage{picture}
\usepackage[linesnumbered,lined,commentsnumbered]{algorithm2e}
\usetikzlibrary{shapes}
\usetikzlibrary{plotmarks}
\usepackage{floatrow}
\newfloatcommand{capbtabbox}{table}[][\FBwidth]

\newcommand{\blind}{1}

\usepackage{titlesec}
\titleformat*{\section}{\large\bfseries}
\titleformat*{\subsection}{\large\bfseries}
\titleformat*{\subsubsection}{\normalsize\bfseries}

\renewcommand\footnotemark{}

\newcommand{\bbeta}{\boldsymbol{\beta}}

\newcommand{\bpsi}{\boldsymbol{\psi}}

\newcommand{\bx}{\mathbf{x}}

\newcommand{\by}{\mathbf{y}}

\newcommand{\boldf}{\boldsymbol{f}}

\newcommand{\bc}{\mathbf{c}}
\newcommand{\be}{\mathbf{e}}

\newcommand{\T}{\mathrm{T}}

\newcommand{\given}{\,|\,}

\begin{document}

\def\spacingset#1{\renewcommand{\baselinestretch}%
{#1}\small\normalsize} \spacingset{1}

%%%%%%%%%%%%%%%%%%%%%%%%%%%%%%%%%%%%%%%%%%%%%%%%%%%%%%%%%%%%%%%%%%%%%%%%%%%%%%

\if1\blind
{
  \title{\bf \Large Bayesian design of experiments for generalised linear models and dimensional analysis with industrial and scientific application
  \vspace{0cm}
  }
  \author{   
    \large David C. Woods$^\dagger$\footnote{Contact: David Woods; \texttt{D.Woods@southampton.ac.uk}; Southampton Statistical Sciences Research Institute, University of Southampton, Southampton SO17 1BJ UK}, Antony M. Overstall$^\ddagger$, Maria Adamou$^\dagger$ and Timothy W. Waite$^\star$ \\[1ex]
    $^\dagger$University of Southampton, Southampton, UK \\
    $^\ddagger$University of Glasgow, Glasgow, UK \\
    $^\star$University of Manchester, Manchester, UK 
    \large }
    \date{\vspace{-1.2cm}}
  \maketitle
} \fi

\bigskip
\begin{abstract}
The design of an experiment can be always be considered at least implicitly Bayesian, with prior knowledge used informally to aid decisions such as the variables to be studied and the choice of a plausible relationship between the explanatory variables and measured responses. Bayesian methods allow uncertainty in these decisions to be incorporated into design selection through prior distributions that encapsulate information available from scientific knowledge or previous experimentation. Further, a design may be explicitly tailored to the aim of the experiment through a decision-theoretic approach using an appropriate loss function. We review the area of decision-theoretic Bayesian design, with particular emphasis on recent advances in computational methods. For many problems arising in industry and science, experiments result in a discrete response that is well described by a member of the class of generalised linear models. Bayesian design for such nonlinear models is often seen as impractical as the expected loss is analytically intractable and numerical approximations are usually computationally expensive. We describe how Gaussian process emulation, commonly used in computer experiments, can play an important role in facilitating Bayesian design for realistic problems. A main focus is the combination of Gaussian process regression to approximate the expected loss with cyclic descent (coordinate exchange) optimisation algorithms to allow optimal designs to be found for previously infeasible problems. We also present the first optimal design results for statistical models formed from dimensional analysis, a methodology widely employed in the engineering and physical sciences to produce parsimonious and interpretable models. Using the famous paper helicopter experiment, we show the potential for the combination of Bayesian design, generalised linear models and dimensional analysis to produce small but informative experiments. \\[1ex]
This paper will appear in Quality Engineering \\ (\texttt{http://www.tandfonline.com/doi/full/10.1080/08982112.2016.1246045}).
\end{abstract}

\noindent%

\section{Introduction}
\label{sec:intro}

Design of experiments is an ``a priori'' activity, taking place before data has been collected, and hence the Bayesian paradigm is a particularly appropriate approach to take. Bayesian methods allow available prior information on the model to be incorporated into both the design of the experiment and the analysis of the resulting data, and produce posterior distributions that are interpretable by scientists. They also reduce reliance on unrealistic assumptions and asymptotic results that may be inappropriate for small- to medium-sized experiments. The Bayesian approach to design enables realistic and coherent accounting for the substantial model and parameter uncertainties that usually exist before an experiment is performed and it is also a natural framework for sequential inference and design.

An important problem where Bayesian methods can have substantial impact is optimal design for nonlinear modelling, which relies on some prior information being available about the unknown values of the model parameters (see \citealp{ADT2007}, ch.~17). A Bayesian approach relaxes the requirement of locally optimal design criteria to specify particular values of the parameters. Fully Bayesian design, predicated on using the posterior distributions for inference as outlined below, is also less reliant on the asymptotic assumptions that underpin most classical design for nonlinear models.

A decision-theoretic Bayesian optimal design is found through minimisation of the expectation of a loss function that is chosen to encapsulate the aims of the experiment. Suppose that we require a design for $q$ variables in $n$ points, with the $i$th point defined as $\bx_i = (x_{i1},\ldots,x_{iq})^\T\in\mathcal{X}\subset\mathbb{R}^q$. The choice of design space $\mathcal{X}$ determines the quality of design that can be found and should be chosen in conjunction with the statistical model. For a linear model composed of only first-order terms and interactions, a discrete design space containing only the corner points of, for example, the unit cube will suffice for finding an optimal design. For the more complex nonlinear models considered in this paper, a continuous design space is needed to allow an optimal design to be found.

Let $l(\xi,\by,\bpsi)$ be the loss function for a design $\xi=\{\bx_1,\ldots,\bx_n\}\in\mathcal{X}^n$ producing data $\by = (y_1,\ldots,y_n)^\T\in\mathcal{Y}$. Assume a statistical model defined via likelihood $p(\by\given\bpsi)$, with parameters $\bpsi\in\Psi$ having prior density $p(\bpsi)$. The vector $\bpsi$ may include parameters defining both the mean and variance of $\by$. Then an optimal design $\xi^\star$ is defined as

\begin{equation}\label{eq:exput}
\xi^\star = \mbox{argmin}_{\xi\in\mathcal{X}^n} \int_{\mathcal{Y}} \int_{\Psi} l(\xi,\by,\bpsi) p(\bpsi,\by\given\xi)\,d\bpsi\,d\by\,.
\end{equation}
For further details, see the landmark review paper of \citet{CV1995}.
%\noindent If the aim of the experiment is prediction, we define $l_p(\xi,\tilde{\by},\by)$ as the loss of the prediction $\tilde{\by}$, and hence the loss function in~(\ref{exput}) is given by $l(\xi,\by,\bpsi) = \int l_p(\xi,\tilde{\by},\by)p(\tilde{\by}\given\bpsi,\by)\,d\tilde{\by}$. 

There are a number of challenges in calculating the expected loss in~(\ref{eq:exput}): 
\begin{enumerate}
\item[(a)] the evaluation of $l$ itself is potentially non-trivial, as it may depend on the posterior distribution and only be available numerically; 
\item[(b)] the integrals in~(\ref{eq:exput}) may be very high dimensional, and are unlikely to be analytically tractable; 
\item[(c)] evaluation of the joint density $p(\bpsi,\by\given\xi) = p(\by \given \bpsi,\xi)p(\bpsi)$ may be complicated by the computational expense of calculating the likelihood $p(\by \given \bpsi,\xi)$ for complex models. For example, hierarchical nonlinear models, perhaps resulting from restricted randomisation in the experiment, lead to analytically intractable likelihoods. Alternatively, the likelihood may result from the evaluation of a complex numerical model and hence not be available in closed-form.
%\item[(d)] the joint probability model must take account of any hierarchies in the experiment structure (e.g. through the application of linear or nonlinear mixed models).
\end{enumerate}

Common choices for $l$ include (i) the self-information loss $p(\bpsi)-\log p(\bpsi\given\by,\xi)$, and (ii) the squared-error loss between (a function of) $\bpsi$ and its posterior expectation. If the prior density does not depend on the design, minimisation of the expected self-information loss is equivalent to maximisation of the expected Kullback-Leibler(KL) divergence between the prior and posterior distributions (see \citealp{mackay2003}, ch.~2, and \citealp{SW2000}). For some experiments, bespoke loss functions may be required. For example, it may be necessary or desirable to incorporate the cost of each run of the experiment. We demonstrate results using both expected self-information loss (SIL), defined as
\begin{equation}\label{eq:NSIG}
	\Phi(\xi)_{\mbox{\footnotesize SIL}} = \left.\int_{\mathcal{Y}}\int_{\Psi} \left[\log p(\bpsi)-\log p(\bpsi \given\by,\xi)\right]p(\bpsi,\by\given\xi)\right.\,d\bpsi\, d\by\,,
\end{equation}
and the expected squared-error loss (SEL) for the prediction of the mean response $\mu(\bx)$,
\begin{equation}\label{eq:SEL}
	\Phi(\xi)_{\mbox{\footnotesize SEL}} = \left.\int_{\mathcal{Y}}\int_{\Psi} \int_\mathcal{X} \left\{\mu(\bx) - E[\mu(\bx) \given \by, \xi]\right\}^2p(\bpsi,\by\given\xi)\right.\,d\bx\,d\bpsi\, d\by\,.
\end{equation}
We refer to a design minimising~\eqref{eq:NSIG} or~\eqref{eq:SEL} as SIL-optimal or SEL-optimal, respectively.

Until very recently, optimal Bayesian design has not evolved far from the methods reviewed by \citet{CV1995}. Development and application of methods for Bayesian design have lagged behind the progress made in inference and modelling due to the additional complexity introduced by the need to integrate over the (as yet) unobserved responses, in addition to unknown model parameters. Hence, methodology has been restricted to simple models and fully sequential, one-point-at-a-time, procedures \citep{RDMP2015}.

In this paper, we focus on experiments for multi-variable generalised linear models and present new results for optimal design for dimensional analysis. In Section~\ref{sec:glms} we introduce the generalised linear models for which optimal designs will be sought and use Box's helicopter experiment to introduce and demonstrate the concepts of dimensional analysis. In Section~\ref{sec:approaches}, we then review the main approaches to overcoming challenges (a)-(c) that have been proposed in the literature. A major focus is the approximate coordinate exchange methodology proposed by \citet{OW2015}. In Section~\ref{sec:results} we use approximate coordinate exchange to find Bayesian optimal designs for three examples, including the helicopter experiment. We finish with a short discussion in Section~\ref{sec:disc}, highlighting issues that may hinder the adoption of Bayesian design in practice, and propose some potential remedies.

\section{Experiments with generalized linear models}\label{sec:glms}

Generalised linear models (GLMs; \citealp{MN1989}) are an important class of models for scientific and industrial experiments whose response cannot be well described by a normal-theory linear model (see \citealp{MMVR2010}). In addition to standard linear regression, the class of GLMs includes models for binary and count data. A GLM has three components:
\begin{enumerate}
\item A distribution for the univariate response $y(\bx)$ taken from the exponential family.
\item A linear predictor $\eta(\bx) = \boldf^\T(\bx)\bbeta$, with $p$-vector $\boldf(\bx)$ holding known functions of the explanatory variables and $\bbeta$ a $p$-vector of unknown model parameters.
\item A link function $g(\mu(\bx)) = \eta(\bx)$ relating $\bx$ to the mean $E\{y(\bx)\} = \mu(\bx)$.
\end{enumerate} 

The variance of $y(\bx)$ takes the form $\mbox{Var}\{y(\bx)\} = \phi V\{\mu(\bx)\}$, with $\phi>0$ a dispersion parameter and the form of the function $V$ dependent on the selected exponential family distribution.

\citet{AW2015} reviewed the upsurge in the development of design methodology for GLMs that has taken place over the last 10 years or so. Key to frequentist, and much Bayesian, optimal design for GLMs is the Fisher information matrix for $\bbeta$, which for an $n$-run experiment takes the form
\begin{equation}\label{eq:info}
M(\bbeta;\,\xi) = X^\T WX\,,
\end{equation} 
with $X$ an $n\times p$ model matrix with $i$th row equal to $\boldf^\T(\bx_i)$ and $W$ an $n\times n$ diagonal matrix with $i$th entry
$$
w(\bx_i) = \left[\mbox{Var}\{y(\bx_i)\}\right]^{-1}\left(\dfrac{d\mu(\bx_i)}{d\eta(\bx_i)}\right)^2\,,\quad i=1,\ldots,n\,.
$$

In general, the information matrix depends on the values of the unknown model parameters $\bbeta$ through the matrix $W$. A notable exception is for the linear regression model. 

Below we discuss two classes of models: (i) GLMs for discrete responses; and (ii) a GLM for a continuous response with a linear predictor that incorporates physical principles via dimensional analysis. 

\subsection{Experiments with discrete responses}\label{sec:discrete}

Perhaps the most common examples of optimal design for GLMs involve discrete responses, for example binary, binomial or count data. \citet{WLER2006} and \citet{WV2011} described examples from chemistry, food technology and engineering with binary (success/fail) responses. In particular, the potato-packing experiment from \citet{WLER2006} involved measuring the formation, or not, of moisture in a protected atmosphere package. The treatments consisted of the settings of three variables: vitamin concentration in the pre-packing dip and the levels of two gases in the atmosphere. A suitable GLM here might be a logistic regression. Let $y(\bx_i)\sim\mbox{Bernoulli}\{\rho(\bx_i)\}$ be the response from the $i$th run of the experiment, with variable settings $\bx_i$ and
\begin{equation}\label{eq:logmodel}
\log\left( \frac{\rho(\bx_i)}{1-\rho(\bx_i)} \right) = \beta_0 + \sum_{j=1}^3\beta_jx_{ij} + \sum_{j=1}^3\sum_{k\ge j}^3\beta_{jk}x_{ij}x_{ik}\,,
\end{equation}
where $\beta_0,\ldots,\beta_3, \beta_{11}, \beta_{12}, \ldots, \beta_{33}$ are unknown parameters to be estimated. Here, $\mu(\bx_i) = \rho(\bx_i)$ and the variance function is given by $V\{\mu(\bx_i)\} = \rho(\bx_i)[1-\rho(\bx_i)]$ with $\phi = 1$. To illustrate some Bayesian design concepts, \citet{AW2015} assumed the following independent prior distributions for the parameters:
\begin{equation}\label{eq:logisticpriors}
\beta_1, \beta_2 \sim\mbox{U}(2,6)\,,\quad \beta_0, \beta_3, \beta_{jk}\sim\mbox{U}(-2,2) \mbox{ for } j,k=1,2,3\,.
\end{equation}

For multi-variable experiments, most theoretical progress on optimal design has been made for Poisson distributed responses, see for example \citet{RWLE2009}. Poisson regression is often employed in industrial experiments counting numbers of defects (\citealp{WH2009}, ch.~14) or in environmental and biological experiments where the response is the count of animal numbers or cell growth. Let $y(\bx_i)\sim\mbox{Poisson}\{\mu(\bx_i)\}$, with $V\{\mu(\bx_i)\} = \mu(\bx_i)$ and $\phi=1$. \citet{ME2012} and \citet{AW2015} presented theoretical constructions of optimal designs robust to the values of the model parameters for log-linear models with linear predictors of the form
\begin{equation}\label{eq:poislp}
\log\left\{\mu(\bx_i)\right\} = \beta_0 + \sum_{j=1}^q\beta_jx_{ij}\,,\quad i=1,\ldots,n\,.
\end{equation}
The latter authors illustrated the construction methods for experiments with $q=5$ variables with uniform prior distributions assumed for each $\beta_j$, 
\begin{equation}\label{eq:poisprior}
\mbox{U}(1, 1+\alpha) \mbox{ for } j=1, 3, 5 \mbox{ and } \mbox{U}(-1-\alpha, -1) \mbox{ for } j=2,4\,, 
\end{equation}
with $\alpha>0$. The intercept $\beta_0=0$ was assumed known.  

In Section~\ref{sec:discretedesign}, we find, assess and compare Bayesian optimal designs for both the logistic and log-linear models.

\subsection{Dimensional analysis}\label{sec:da}

%\begin{table}
%\begin{center}
%\begin{tabular}{ll}
%Variable/parameter (symbol) & Range/setting \& units \\
%\hline
%Flight time ($T$) & s \\ \\
%Rotor width ($w$) & $\in [0.03,0.09]\,$m \\
%Rotor length ($r$) & $\in [0.07, 0.12]\,$m \\
%Tail length ($l$) & $\in [0.07, 0.12]\,$m \\ \\
%Mass ($m$) & kg \\
%Drop height ($h$) & Fixed at $2\,$~m \\
%Acceleration due to gravity ($g$) & Fixed at $9.80665\,$ms$^{-2}$ \\
%Air density ($\rho$) & Fixed at $1.20412\,$kgm$^{-3}$ \\
%Paper density ($D$) & Fixed at $0.120\,$kgm$^{-2}$ \\
%Body length ($b$) & Fixed at $0.025\,$m \\
%Tail width ($d$) & Fixed at $0.05\,$m
%\end{tabular}
%\end{center}
%\caption{\label{tab:paperhelidim}Variable ranges and physical parameter settings (in SI units) for the paper helicopter experiment. Flight time is the measured response.}
%\end{table}

Dimensional analysis (DA) is a methodology commonly used by engineers and physical scientists to produce parsimonious and dimensionally consistent models (\citealp{Sonin2001}). A base set of dimensionless variables are identified via (nonlinear) transformations of the explanatory variables and related to a similarly transformed response variable via a, typically nonlinear, regression model. The model thus formed will satisfy Buckingham's $\Pi$ theorem \citep{Buckingham1914,Buckingham1915a,Buckingham1915b}, which states that physically meaningful relationships must be dimensionally homogeneous. In addition to performing a priori dimension reduction of the input variables, DA provides the possibility of obtaining models that are scale-free and hence, for example, applicable to a range of manufacturing processing scales from lab to production.

Reviews of DA from a statistician's perspective are provided by the recent papers from \citet{ANAC2013} and \citet{SDLN2014}. The latter authors applied DA to Box's paper helicopter experiment \citep{BL1999}, and we will use this example to demonstrate the potential of the combination of DA, GLMs and Bayesian design.

\begin{figure}
\begin{floatrow}
\ffigbox[\FBwidth]{%
\includegraphics[scale = .5]{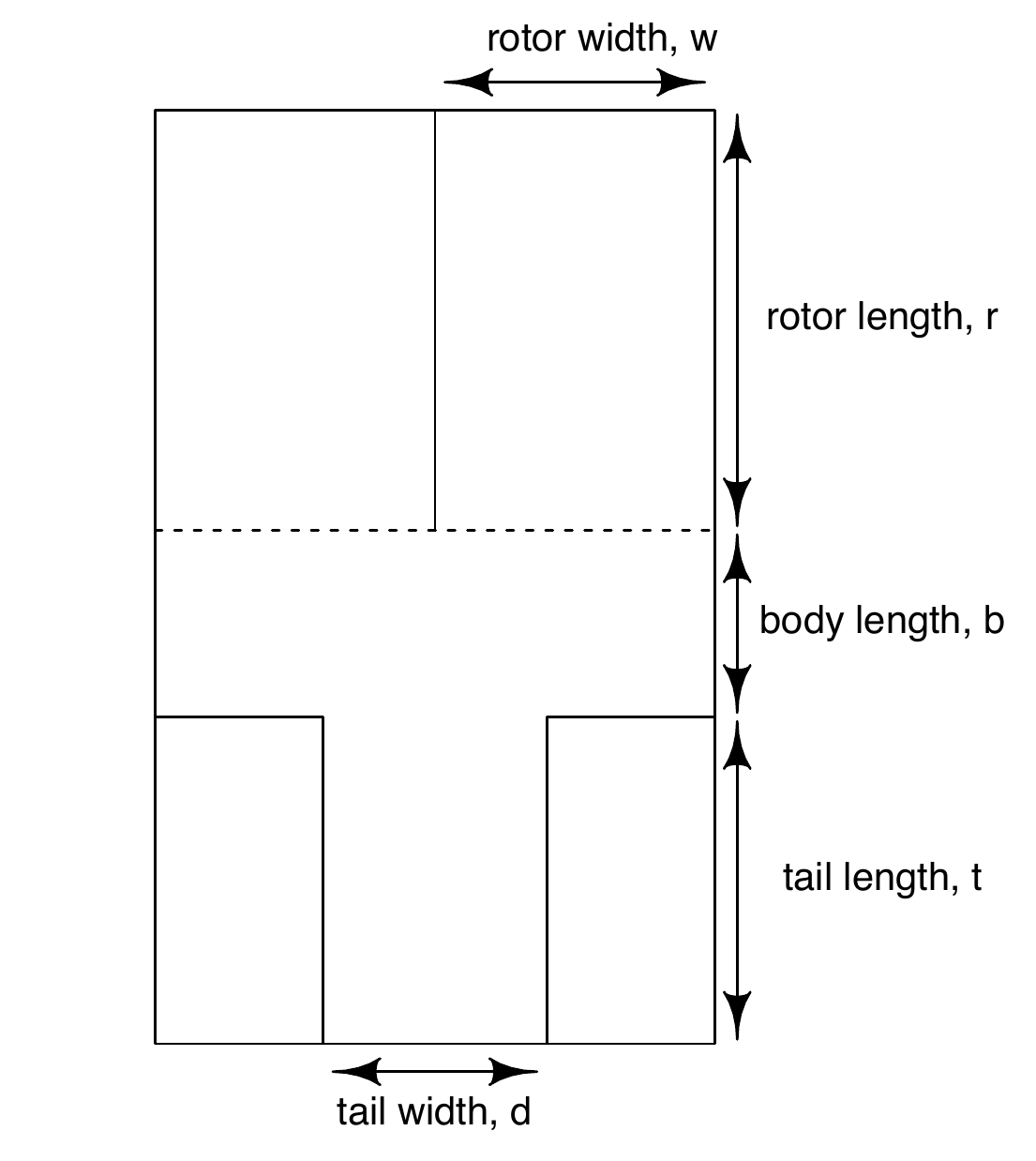} 
}{%
\caption{\label{fig:paperheli}Paper helicopter pattern.}%
}
\capbtabbox{%
 \begin{tabular}{ll}
Variable/parameter (symbol) & Range/setting \& units \\
\hline
Flight time ($T$) & s \\ \\
Rotor width ($w$) & $\in [0.03,0.09]\,$m \\
Rotor length ($r$) & $\in [0.07, 0.12]\,$m \\
Tail length ($t$) & $\in [0.07, 0.12]\,$m \\ \\
Mass ($m$) & kg \\
Drop height ($h$) & Fixed at $2\,$~m \\
Acceleration due to gravity ($g$) & Fixed at $9.80665\,$ms$^{-2}$ \\
Air density ($\rho$) & Fixed at $1.20412\,$kgm$^{-3}$ \\
Paper density ($D$) & Fixed at $0.120\,$kgm$^{-2}$ \\
Body length ($b$) & Fixed at $0.025\,$m \\
Tail width ($d$) & Fixed at $0.05\,$m
\end{tabular}

}{%
  \caption{\label{tab:paperhelidim}Variable ranges and physical parameter settings (in SI units) for the paper helicopter experiment. Flight time is the measured response.}%
}
\end{floatrow}
\end{figure}

We use a standard paper helicopter pattern and consider three independent variables: rotor width, rotor length and tail length, see Figure~\ref{fig:paperheli}. The ranges of these three variables, and other dimensions, are taken from \texttt{http://www.paperhelicopterexperiment.com} and are given in Table~\ref{tab:paperhelidim}, along with settings of other physical parameters.

\citet{SDLN2014} identified the base quantities 
\begin{equation*}
\Pi_0 = \dfrac{h}{T\sqrt{gr}}\,,\qquad \Pi_1 = \dfrac{m}{\rho r^3}\,,
\end{equation*}
and fitted a regression model for flight time on the log scale with
\begin{eqnarray}
\log E(T) = \log\theta_0 + \log\dfrac{h}{\sqrt{gr}} + \theta_1\log\left[\dfrac{\rho r^3}{m}\right] \label{eq:powerlaw}\\
\Rightarrow E(T) = \dfrac{\theta_0h}{\sqrt{gr}}\left[\dfrac{\rho r^3}{m}\right]^{\theta_1}\,.\nonumber 
\end{eqnarray}
Equation~\eqref{eq:powerlaw} follows a power law and is consistent with Bridgman's principle of absolute significance of relative magnitude \citep{Bridgman1931}, another fundamental theorem of DA.

In common with \citet{SDLN2014}, we consider experiments with $n=4$ runs but adapt these authors' study in two ways:
\begin{enumerate}
\item We assume the density of the paper is fixed, and hence helicopter mass ($m$) is a varying function of the three controllable variables rotor width, rotor length and tail length. Relabeling $T = y(\bx)$ and $r = x_1$, $w = x_2$ and $t = x_3$ for consistency, equation~\eqref{eq:powerlaw} then becomes
\begin{equation}\label{eq:helilinpred}
\log E\{y(\bx)\} = \log \mu(\bx) = \log\theta_0 + \log\dfrac{h}{\sqrt{gx_1}} + \theta_1\log\left\{\dfrac{\rho x_1^3}{D[2x_2(x_1+b) + x_3d]}\right\}\,.\\
\end{equation}
\item Rather than assume additive normal errors on the log scale, we model flight time as a Gamma distributed random variable and assume a GLM with log link and linear predictor~\eqref{eq:helilinpred}. Hence $V\{\mu(\bx)\} = \phi\mu(\bx)^2$ and $\phi$ is assumed unknown. The second term on the right-hand side of~\eqref{eq:helilinpred} is treated as an offset, and hence there are three parameters, $\log\theta_0$, $\theta_1$ and $\phi$, that require estimation. 
\end{enumerate}

We assume prior distributions 
\begin{equation}\label{eq:daprior}
\log\theta_0 \sim N(0.102, 0.0625)\,,\quad\theta_1\sim N(0.460, 0.0625)\,,\quad\phi \sim U(0.75, 1.25)\,.
\end{equation}
The prior means of the physical parameters are equal to estimates from \citet{SDLN2014}; we set the prior variances equal to 2.5 times the estimated variances from \citet{SDLN2014} to obtain more diffuse prior distributions. For the dispersion parameter, our choice of prior leads to the variance of $y(\bx)$ being between 75\% and 125\% of the value of $\mu(\bx)^2$.

In Section~\ref{sec:daresults}, we find, assess and compare Bayesian optimal designs for this Gamma regression model.

\section{A review of approaches to Bayesian design}
\label{sec:approaches}

In this section, we provide an overview of some of the most common approaches to Bayesian design in the literature. We focus on (i) analytical and computational approximations to the expected loss in~\eqref{eq:exput}, and (ii) optimization methods for multi-variable experiments.

\subsection{Asymptotic approximations}\label{sec:asym} 
For experiments with large $n$, the inverse of the expected Fisher information matrix $M(\bpsi;\,\xi)$ is an asymptotic approximation to the posterior variance-covariance matrix of the parameters $\bpsi$. Use of this approximation leads to pseudo-Bayesian ``alphabetic'' optimality criteria (\citealp{ADT2007}, ch.~10). For example, under pseudo-Bayesian $D$-optimality, a design is selected to minimise 
\begin{equation}\label{eq:dopt}
\Phi_D(\xi) = \int_{\Psi} - \log \left| M\left(\bpsi;\,\xi \right) \right| \pi(\bpsi)\,d\bpsi\,.
\end{equation}
%and $A$-optimality minimises 
%$$\Psi(\xi) = \int_{\Theta} \mbox{tr}\left\{M^{-1}(\btheta;\,\xi)\right\}\pi(\btheta)\,\mathrm{d}\btheta\,.$$
The integral with respect to $\bpsi$ is usually of low dimension and amenable to deterministic approximation. Such an approximation to the objective function can then be minimised using a conditional algorithm such as point or coordinate exchange; see, for, example, \citet{GJS2009}. For a point prior density on $\bpsi$, which is equivalent to assuming known parameter values, minimisation of~\eqref{eq:dopt} leads to a locally $D$-optimal design.

\subsection{Simulation-based optimisation}\label{sim} 
In general, the expected loss can be approximated via Monte Carlo integration (e.g. \citealp{Gentle2003}, ch.~7) as
$$
\tilde{\Phi}(\xi) = \frac{1}{B}\sum_{k=1}^B 
l\left(\xi, \by_k, \bpsi_k\right)\,,
$$ 
with $(\bpsi_k,\by_k)\sim p(\bpsi,\by\given\xi)$ a random sample drawn from the joint distribution of parameters $\bpsi$ and responses $\by$. Typically, this is obtained by sampling $\bpsi$ from the prior density $p(\bpsi)$ and then sampling $\by$ from the conditional density $p(\by\given \bpsi,\xi)$. The loss $l(\xi, \by_k,\bpsi_k)$ often itself requires numerical approximation, necessitating a nested, double-loop, Monte Carlo simulation; see \citet{Ryan2003}. Direct optimisation of this approximation requires large $B$ to generate a suitably precise approximation to the objective function and/or expensive stochastic algorithms (e.g. simulated annealing or genetic algorithms), see for example, \citet{HMRW2001} and \cite{HM2013} who employed polynomial chaos approximations to facilitate sampling from $p(\bpsi,\by\given\xi)$. Alternatively, the optimisation can be embedded within a Markov chain simulation scheme, and an optimal design identified by sampling from an artificial joint distribution for the design, model parameters and data, and then finding the mode of the marginal distribution of the design (\citealp{Muller1999}, \citealp{MSD2004} and \citealp{ABPR2006}). Typically, an annealing step is employed to enable easier identification of the optimal design. This approach is most effective for small experiments (few variables and runs). Recent extensions to this algorithm have allowed designs to be found for (i) models with intractable likelihoods using Approximate Bayesian Computation (see \citealp{DP2013}, and also \citealp{HMW2013} for other ABC design methods) and (ii) dynamic models with numerous sampling times using dimension reduction \citep{RDTP2014}, importance sampling and Laplace approximations \citep{RDP2015}.  

\subsection{Sequential design}\label{sec:seq} 
Most experiments are part of a sequence, where a Bayesian approach, with sequential updating from prior to posterior distributions, is natural. For point-sequential designs, with one point at a time added to the design, approximation of the expected loss is greatly simplified by the resulting reduction in the dimension of the integral. Recent methods have been suggested for estimation of, and discrimination between, nonlinear models, see \citet{DMP2013,DMP2014}. A growing area is Bayesian optimisation of expensive black-box functions (e.g.~in computer experiments), using Gaussian process surrogates to reduce the number of required function evaluations, following the seminal work of \citet{JSW1998}. The computational efficiency of sequential design can be greatly aided through the use of sequential Monte Carlo for the necessary inference \citep[see][]{GP2011}.

\subsection{Smoothing-based optimisation}\label{sec:smooth} 

%\begin{figure}	
%\begin{center}
%\includegraphics[scale=0.5]{figs/comp1d.pdf}
%\end{center}
%\caption{\label{fig:comp1d}Compartmental example, $n=2$ with the first design point fixed at time $t_1=5$ (vertical line) and smoothed approximate NSIG for 10 values of the second design point $t_2$.}
%\end{figure}	

Smoothing-based design methods \citep{MP1996} evaluate a computationally expensive, typically Monte Carlo, approximation to the expected loss~\eqref{eq:exput} for a limited number of designs and then smooth these approximated losses to generate a surrogate function which can then be optimised in place of the true expected loss. Recent research includes (i) extension of such methods to employ Gaussian process smoothing and Bayesian optimisation \citep{WWAH} and (ii) use of surrogates to enable Bayesian $D$-optimal design for generalised linear mixed models \citep{WW2015}. %See also \cite{HM2013,HM2014}.

%We demonstrate the \citet{MP1996} smoothing method by finding a simple design for compartmental model~\eqref{eq:compmodel}. We fix the first point in a design with $n=2$ at $t_1=5$, and approximate the NSIG for 10 designs with different values of the second point $t_2$. Figure~\ref{fig:comp1d} shows the smoothed NSIG constructed from these 10 designs, as a function of $t_2$. Such a smoother may be used to find an optimal value of $t_2$, which is about $t_2=17$ in this example. 

%\textbf{Smoothing the Bayes risk:} Assume a stochastic (Monte Carlo) approximatation, $\tilde{\Psi}$, of the Bayes risk is available. M{\"u}ller and Parmigiani (1995) found optimal designs via curve fitting: (1) evaluate $\tilde{\Psi}$ for designs $\xi_1.\ldots,\xi_Q$; (2) fit a model $\hat{\Psi}(\xi)$ to $\{\tilde{\Psi}(\xi_i),\xi_i)$; and (3) find $\xi$ that minimises $\hat{\Psi}(\xi)$. 

A key challenge for the application of smoothing-based methods to design problems with large numbers of runs or variables is the high-dimensional smoothing that is required. An extension that addresses this challenge using conditional smoothing and optimisation is described in the next section.

\subsection{Approximate coordinate exchange}\label{sec:ACE}
\citet{OW2015} proposed the first general purpose methodology for high-dimensional, multi-variable Bayesian design that does not rely on normal approximations to the posterior density. Their approximate coordinate exchange (ACE) algorithm is a conditional optimisation algorithm that makes use of surrogates, or emulators, for the expected loss as a function of a single design coordinate (i.e.~the value of a single variable for a particular run).

In the field of computer experiments \citep{sacks}, an emulator is a statistical model built to approximate the output from an expensive computer model or code. The most common emulator is a Gaussian process model (GP; see, for example, \citealp{RW2006}), which non-parametrically smooths or interpolates the computer model output. In the ACE algorithm, GP models are used to smooth the Monte Carlo approximation to the expected loss.

Algorithm~\ref{alg:ace} gives the basic steps of ACE. Let $\tilde{\Phi}_{ij}(x \given \xi)$ be a Monte Carlo approximation to the expected loss for design $\xi$ with $ij$th coordinate replaced by $x\in\mathcal{X}_j$ $(i=1,\ldots,n;\,j=1,\ldots,q)$. Here, $\mathcal{X}_j$ is the projection of the design space $\mathcal{X}$ onto the $j$th dimension. The algorithm steps through each coordinate of the design, and for each coordinate constructs a one-dimensional emulator, $\hat{\Phi}(x)$, for $\tilde{\Phi}_{ij}(x \given \xi)$. The value of $x$ that minimises this emulator is found, and an accept/reject step performed in order to decide whether to swap the current design coordinate with this proposed coordinate. This accept/reject step is described in Algorithm~\ref{alg:ar}. 

The ACE algorithm would typically be repeated multiple times (perhaps exploiting parallel computing) to avoid local optima. \citet{OW2015} gave more details of the implementation and application of the algorithm, including its combination with a point-exchange algorithm to consolidate clusters of similar design points.

\begin{algorithm}
\SetKwData{Left}{left}\SetKwData{This}{this}\SetKwData{Up}{up}
\SetKwFunction{Union}{Union}\SetKwFunction{FindCompress}{FindCompress}
\SetKwInOut{Input}{Input}\SetKwInOut{Output}{Output}

\Input{Initial (randomly chosen) design $\xi$}
\Output{$\Phi$-optimal design}
\Begin{
\Repeat{$\mathrm{convergence}$}{
\For{$i = 1:n$}{
\For{$j = 1:q$}{
Generate a 1d space-filling design $\zeta_{ij} = \left\{x_{ij}^1,\ldots,x_{ij}^Q\right\}$ in $\mathcal{X}_j\subset\mathbb{R}$\;
\For{$k = 1:Q$}{Evaluate $\tilde{\Phi}_{ij}(x_{ij}^k \,\given\, \xi)$ using Monte Carlo sample size $B$}\;
Construct a 1d emulator $\hat{\Phi}(x)$ via~\eqref{eq:emulator}\;\label{ace:em}
Set the $ij$th coordinate, $x_{ij}$, of $\xi$ equal to $\operatorname{argmin}_{x\in\mathcal{X}_j}\hat{\Phi}(x)$ with probability $p^\star$ obtained from Algorithm~\ref{alg:ar}\;\label{ace:ar}
}
}
}
}
\caption{\label{alg:ace}The approximate coordinate exchange (ACE) algorithm.}
\end{algorithm}

\begin{algorithm}
\SetKwData{Left}{left}\SetKwData{This}{this}\SetKwData{Up}{up}
\SetKwFunction{Union}{Union}\SetKwFunction{FindCompress}{FindCompress}
\SetKwInOut{Input}{Input}\SetKwInOut{Output}{Output}

\Input{Current design $\xi$ and proposed new coordinate $x$}
\Output{Posterior probability $p^\star$ that $\tilde{\Phi}_{ij}(x \given \xi) < \tilde{\Phi}(\xi)$}
\Begin{
Let $\xi_p$ be the design formed by replacing the $ij$th coordinate of $\xi$ with $x$\;
\For{$k =1:\tilde{B}$}{
Sample $\tilde{\bpsi}$ from $\pi(\bpsi)$\;
Sample $\by_1\sim \pi(\by \given \bpsi, \xi_p)$ and $\by_2\sim \pi(\by \given \bpsi, \xi)$\;
Set $L_{1k} = l(\xi_p,\by_1,\tilde{\bpsi})$ and $L_{2k} = l(\xi, \by_2, \tilde{\bpsi})$\;
}
Assume $L_{1k}\sim N(b_1+b_2, a)$ and $L_{2k}\sim N(b_1, a)$\;
Calculate the posterior probability, $p^\star$, that $b_2<0$ using ``data'' $L_{1k}$ and $L_{2k}$\;
}
\caption{\label{alg:ar}Accept/reject step from line~\ref{ace:ar} of the ACE algorithm.}
\end{algorithm}

A GP model is employed in line~\ref{ace:em} of Algorithm~\ref{alg:ace}. The emulator is given by the posterior mean function of the GP
\begin{equation}\label{eq:emulator}
\hat{\Phi}(x) = \hat{\mu}_{ij} + \hat{\sigma}_{ij}\boldsymbol{a}^{\T}(x, \zeta_{ij})A(\zeta_{ij})^{-1}\boldsymbol{z}_{ij}\,,
\end{equation}
with $\hat{\mu}_{ij} = \sum_{k=1}^Q \tilde{\Phi}_{ij}(x_{ij}^k \given \xi)/Q$, $\hat{\sigma}^2_{ij} = \sum_{k=1}^Q \left(\tilde{\Phi}_{ij}(x_{ij}^k \given \xi) - \hat{\mu}_{ij}\right)^2/(Q-1)$, $\boldsymbol{z}_{ij}$ a $Q$-vector having $k$th entry equal to the standardised approximate expected loss
$$
\left(\tilde{\Phi}_{ij}(x_{ij}^k \given \xi) - \hat{\mu}_{ij}\right)/\hat{\sigma}_{ij}\,,
$$ 
and $\zeta_{ij} = \left\{x_{ij}^1,\ldots,x_{ij}^Q\right\}$ being points from a one-dimensional space-filling design in $\mathcal{X}_j$ (see Algorithm~\ref{alg:ace}). Under the common assumption of a squared exponential correlation structure, the correlation between the responses at two points is inversely proportional to the exponential of the squared difference between the points, leading to the $Q$-vector $\boldsymbol{a}$ and $Q \times Q$ matrix $A$ having entries 
$$
\boldsymbol{a}(x, \zeta_{ij})_u = \exp\left\{-\rho(x - x_{ij}^u)^2\right\}\,,\quad A(\zeta_{ij})_{uv} = \exp\left\{-\rho(x_{ij}^u - x_{ij}^v)^2\right\} + \eta\operatorname{I}(u=v)\,,\quad u,v = 1,\ldots,Q\,,
$$
where $\operatorname{I}$ is the indicator function and $\eta>0$ is the nugget. Adding $\eta$ to the diagonal elements of the correlation matrix ensures the emulator will smooth, rather than interpolate, the stochastic Monte Carlo approximation $\tilde{\Phi}$, and improves the numerical stability of the emulator \citep{GL2012}. We estimate $\rho$ and $\eta$ via maximum likelihood.

The minimisation in line~\ref{ace:ar} of Algorithm~\ref{alg:ace} is subject to both Monte Carlo error and emulator error. To remove the emulator error when making the decision whether to accept the exchange, the steps in Algorithm~\ref{alg:ar} are performed using $\tilde{B}$ independent Monte Carlo samples to assess the improvement in the design. Algorithm~\ref{alg:ar} essentially describes a Bayesian $t$-test based on simulated data from the existing and proposed designs (cf \citealp{WZ2006}). If the assumption of normality that underpins this test is invalid, a nonparametric procedure may be used instead. 

Convergence in Algorithm~\ref{alg:ace} is assessed graphically, in a similar spirit to convergence diagnostics for Markov chain Monte Carlo; see \citet{OW2015} for examples.

\section{Bayesian designs for generalized linear models}\label{sec:results}

In this section, we find Bayesian optimal designs for the three GLMs outlined in Section~\ref{sec:glms}. The new designs are found using ACE, with $B=1000$ and $Q=20$ in Algorithm~\ref{alg:ace} and $\tilde{B}=20,000$ in Algorithm~\ref{alg:ar}, unless otherwise stated, and assessed against various competitors from the literature.

\begin{figure}[htbp]
\begin{center}
\begin{tabular}{cc}
(a) & (b) \\
\includegraphics[scale=0.4]{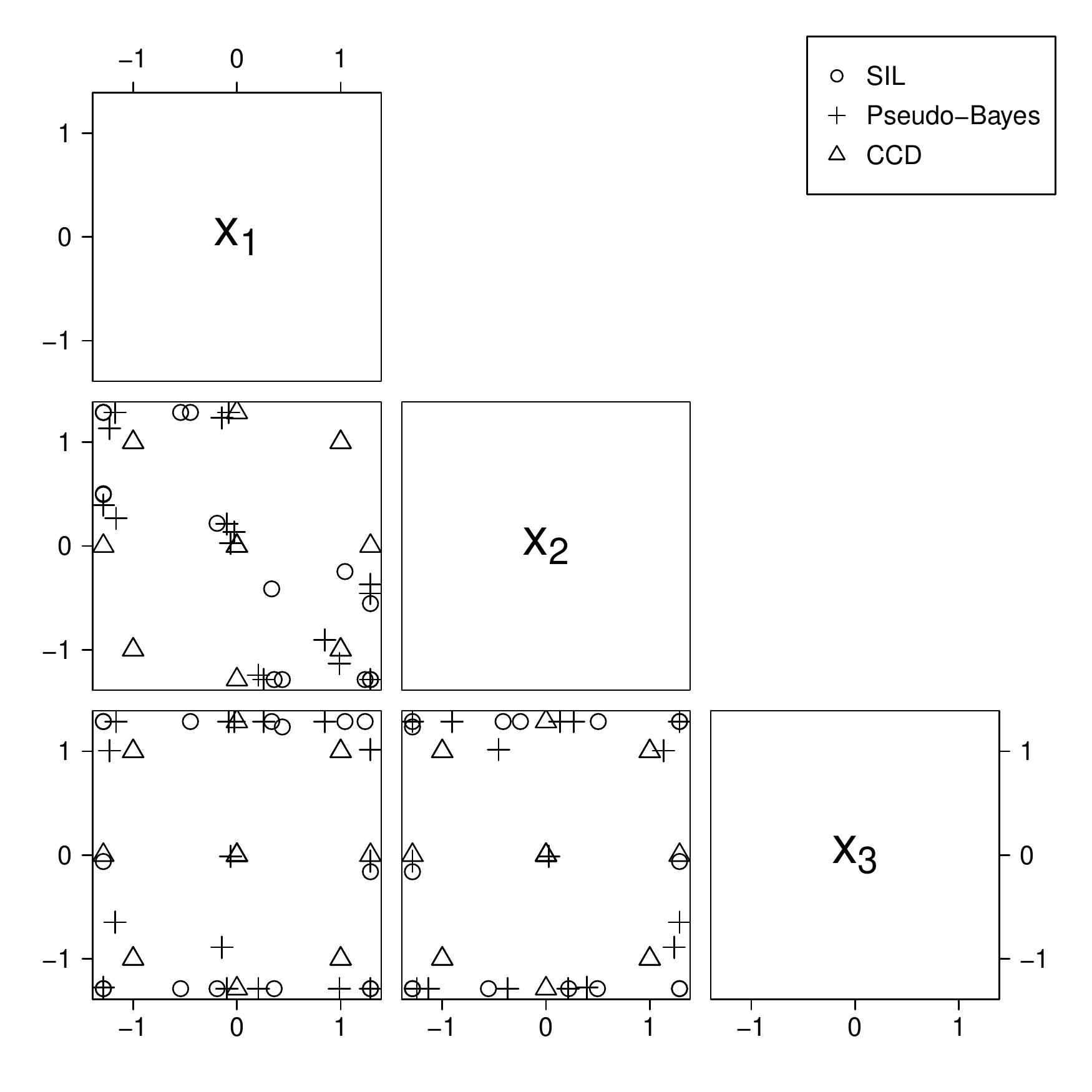} & \includegraphics[scale=0.4]{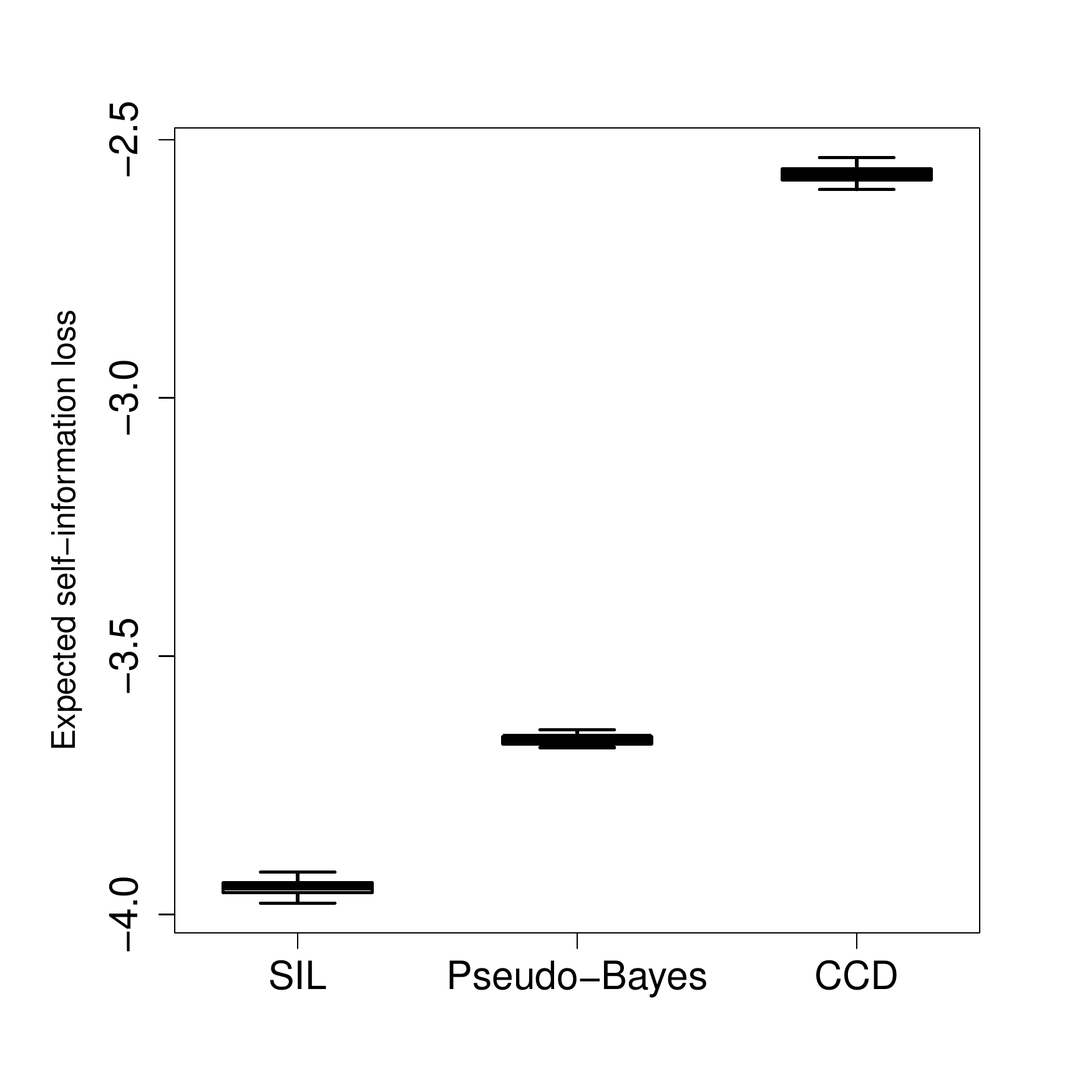}
\end{tabular}
\end{center}
\caption{\label{fig:logistic_designs}Logistic regression example: (a) Two-dimensional projections of the SIL-optimal, pseudo-Bayesian $D$-optimal and central composite designs. (b) Boxplots of 20 Monte Carlo approximations of the negative expected KL divergence~\eqref{eq:NSIG} using $B=20,000$ simulations for the SIL-optimal, pseudo-Bayesian $D$-optimal and central composite designs.}
\end{figure}

\subsection{Optimal design for experiments with discrete response}\label{sec:discretedesign}

We now demonstrate Bayesian design for logistic and Poisson regression. As most previous design work for these models has focussed on $D$-optimality, we find SIL-optimal designs and compare with pseudo-Bayesian $D$-optimal designs. These latter designs are an asymptotic approximation to the SIL-optimal designs (see \citealp{OW2015}).

\subsubsection{Logistic regression}\label{sec:bindesigns}

\begin{figure}[htbp]
\begin{center}
\includegraphics[scale=0.75]{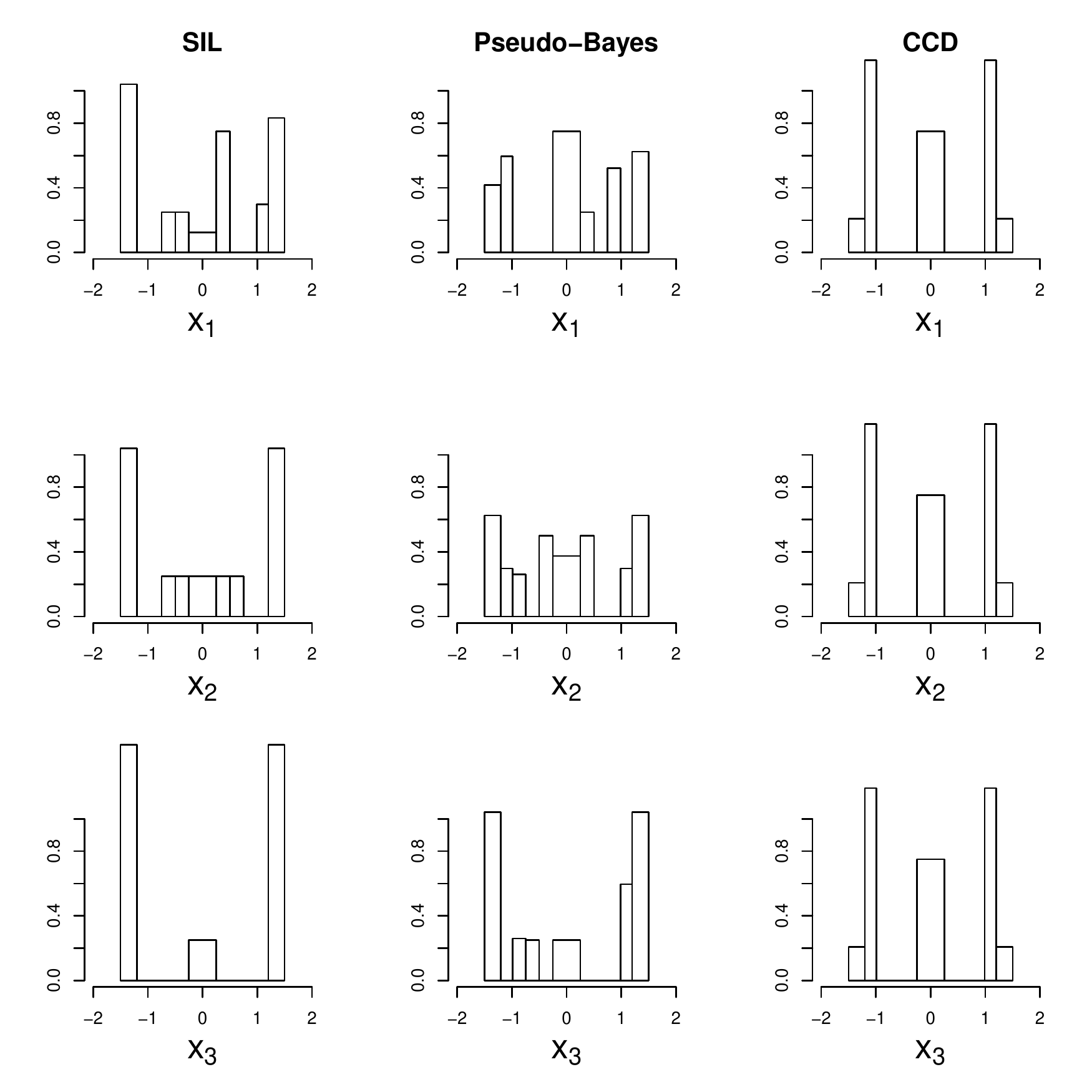}
\end{center}
\caption{\label{fig:logistic_designs_proj}Logistic regression example: one-dimensional projections of the SIL-optimal, pseudo-Bayesian $D$-optimal and central composite designs.}
\end{figure}

We start by finding $n=16$ run SIL-optimal designs for logistic regression, minimising a Monte Carlo approximation to~\eqref{eq:NSIG} and using $Q=10$ in Algorithm~\ref{alg:ace}, with linear predictor~\eqref{eq:logmodel}, $\mathcal{X} = [-1.2872, 1.2872]^3$ and prior distribution~\eqref{eq:logisticpriors}. The choice of design space $\mathcal{X}$ allows comparison to an orthogonal central composite design (CCD) with $n=16$ points. The CCD is a common design for a linear model with a second-order linear predictor (see \citealp{dv1999}, ch.~16). Figure~\ref{fig:logistic_designs}(a) gives two-dimensional projections of the SIL-optimal design, along with the CCD and a pseudo-Bayesian $D$-optimal design with $n=16$ points that minimises a quadrature approximation to~\eqref{eq:dopt}. The main qualitative difference between the SIL-optimal and $D$-optimal designs is the greater concentration of points at the extremes of the design region for the SIL-optimal design, especially for $x_3$. These differences in the distribution of the design points can be clearly seen in Figure~\ref{fig:logistic_designs_proj}, which shows histograms of the one-dimensional projections of the three designs. The concentration of points at the extremes of $x_3$ for both optimal designs is consistent with literature results on locally optimal design with parameter values that are small in absolute value (see \citealp{cox1988}); here, $E(\beta_3) = 0$.

To quantitatively compare the three designs, we performed 20 repeated Monte Carlo approximations of~\eqref{eq:NSIG}, each using $B=20,000$. Boxplots of these results are given in Figure~\ref{fig:logistic_designs}(b). The SIL-optimal design naturally performs best, and has around 10\% lower negative expected KL divergence compared to the pseudo-Bayesian design. Both are substantially better than the central composite design, which of course makes no use of prior information and should only be employed for the estimation of a second-order response surface model which is linear in the unknown parameters.

%The NISG-optimal and pseudo-Bayesian $D$-optimal designs in Figure~\ref{fig:logdesigns} were both found using the ACE algorithm, followed by application of a point exchange algorithm to consolidate close clusters of design points. Both designs are quite different from a factorial design, having multiple values of $x_1$ and, especially, $x_2$ (the two variables that have parameters with prior support not including 0). The pseudo-Bayesian $D$-optimal design also includes some points in the interior of the design region. We compare the performance of these two designs, as well as designs for other values of $n$, by computing 20 independent Monte Carlo approximations to the NSIG, see Figure~\ref{fig:logresults}.
%\clearpage
\subsubsection{Poisson regression}\label{sec:poisdesigns}

\citet{ME2012} and \citet{AW2015} presented analytical construction methods for pseudo-Bayesian optimal designs for log-linear regression with linear predictor~\eqref{eq:poislp}. The resulting designs minimise~\eqref{eq:dopt} amongst the class of minimally supported designs, that is, designs for which the number of distinct design points is equal to the number of parameters in the linear predictor. The construction method uses the algorithm of \citet{RWLE2009}. For $-1 \le x_{ij} \le 1$ $(i=1,\ldots,q+1;\, j=1,\ldots,q)$ and $|E(\beta_j)|>1$, the unreplicated minimally supported pseudo-Bayesian $D$-optimal (MSPBD) design has points $\bx_i = \bc - 2\be_i/E(\beta_i)$ for $i=1,\ldots,q$ and $\bx_{q+1} = \bc$, where $\be_i$ is the $i$th column of the $q\times q$ identity matrix and $\bc = (c_1,\ldots,c_q)^{\T}$ with $c_i = 1$ if $E(\beta_i)>0$ and $c_i=-1$ if $E(\beta_i)<0$. The minimally supported optimal design for $q=5$ variables and prior distribution~\eqref{eq:poisprior} is given in Table~\ref{tab:poisMSdesign}.

Using ACE, we find SIL-optimal designs, again minimising a Monte Carlo approximation to~\eqref{eq:NSIG}, with $n=q+1=6$ design points under prior distribution~\eqref{eq:poisprior}. The designs are given in Tables~\ref{tab:poisBayesdesigns}(a) and~\ref{tab:poisBayesdesigns}(b) for $\alpha = 0.5$ and $\alpha=0.75$, respectively. These designs have the same structure as the MSPBD-optimal design, with each variable only taking two values, one value with $x_{ij} = \pm 1$ and one value with $-1<x_{ij}<1$. Unlike the MSPBD-optimal designs, this latter value is not constant in absolute value across the variables, although it does always have the same sign as the corresponding value in the MSPBD-optimal design.

\begin{table}[htbp]
%MS design
\begin{center}
\begin{tabular}{rrrrrr}
Run & $x_1$ & $x_2$ & $x_3$ & $x_4$ & $x_5$ \\
\hline
1 & $-\gamma$ & -1  & 1 & -1 & 1 \\
2 & 1 &  $\gamma$ & 1 & -1 & 1 \\
3 & 1 & -1 & $-\gamma$ & -1 & 1 \\
4 & 1 & -1 & 1 & $\gamma$ & 1 \\
5 & 1 & -1 & 1 & -1 & $-\gamma$ \\
6 & 1 & -1 & 1 & -1 & 1
\end{tabular}
\end{center}
\caption{\label{tab:poisMSdesign}Log-linear regression example: minimally-supported Pseudo-Bayesian $D$-optimal design under uniform prior distribution~\eqref{eq:poisprior}; $\gamma = 0.6$ for $\alpha = 0.5$ and $\gamma = 0.455$ for $\alpha = 0.75$.}
\end{table}

\begin{table}[htbp]
\begin{center}
\begin{tabular}{cc}
(a) & (b) \\
%a = 0.5
%\begin{center}
\begin{tabular}{rrrrrr}
Run & $x_1$ & $x_2$ & $x_3$ & $x_4$ & $x_5$ \\
\hline
1 & -0.500 & -1 & 1 & -1 & 1 \\
2 & 1 & 0.555 & 1 & -1 & 1 \\
3 & 1 & -1 & -0.309 & -1 & 1 \\
4 & 1 & -1 & 1 & 0.334 & 1 \\
5 & 1 & -1 & 1 & -1 & -0.381 \\
6 & 1 & -1 & 1 & -1 & 1 \\
\end{tabular}
%\end{center}
&
%a = 0.75
%\begin{center}
\begin{tabular}{rrrrrr}
Run & $x_1$ & $x_2$ & $x_3$ & $x_4$ & $x_5$ \\
\hline
1 & -0.220 & -1 & 1 & -1 & 1 \\
2 & 1 & 0.222 & 1 & -1 & 1 \\
3 & 1 & -1 & -0.323 & -1 & 1 \\
4 & 1 & -1 & 1 & 0.110 & 1 \\
5 & 1 & -1 & 1 & -1 & -0.308 \\
6 & 1 & -1 & 1 & -1 & 1 \\
\end{tabular}
%\end{center}
\end{tabular}
\end{center}
\caption{\label{tab:poisBayesdesigns}Log-linear regression example: SIL-optimal designs under uniform prior distribution~\eqref{eq:poisprior} for (a) $\alpha = 0.5$ and (b) $\alpha = 0.75$.}
\end{table}

\begin{figure}[htbp]
\begin{center}
\begin{tabular}{cc}
(a) & (b) \\[-3ex]
\includegraphics[scale=.4]{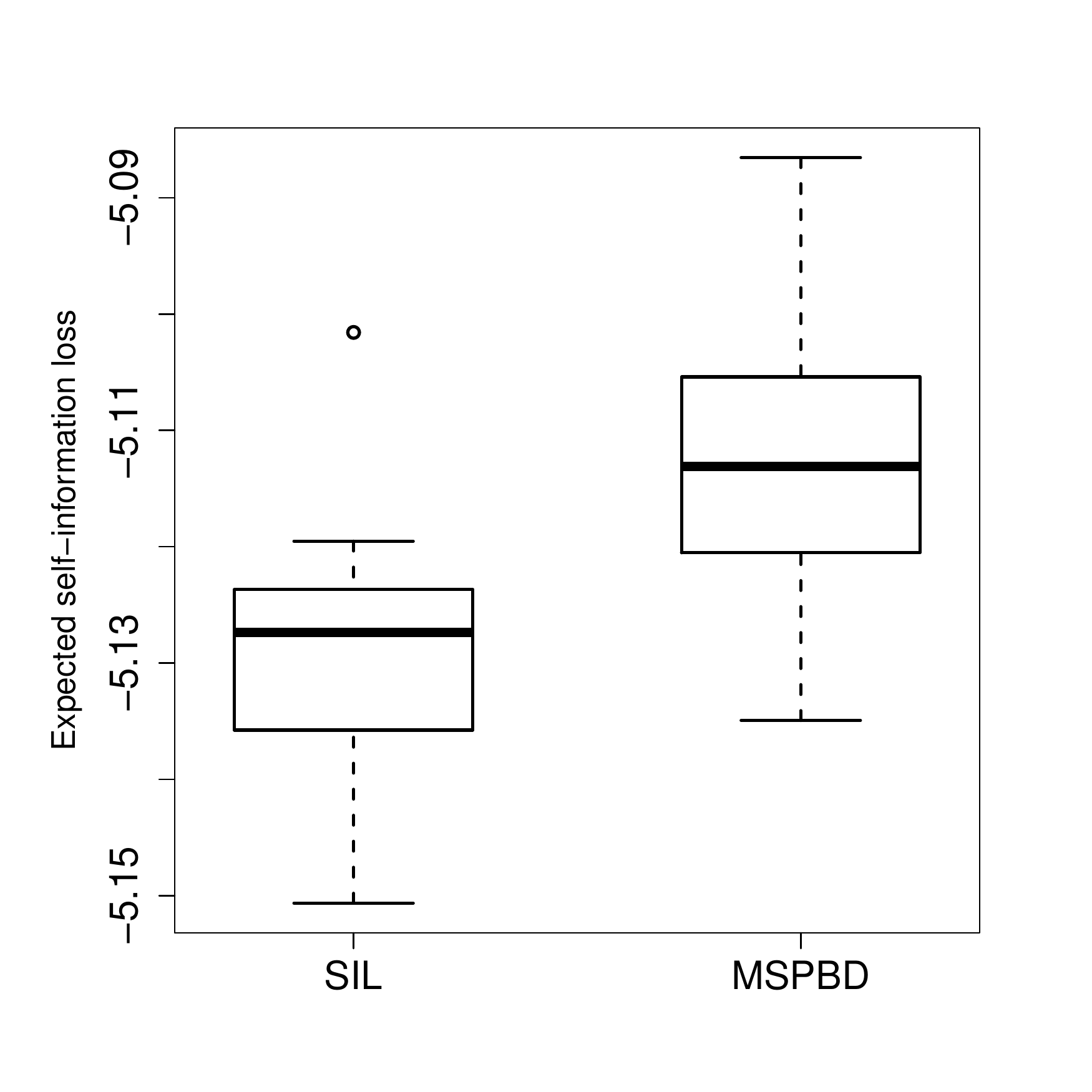} & \includegraphics[scale=.4]{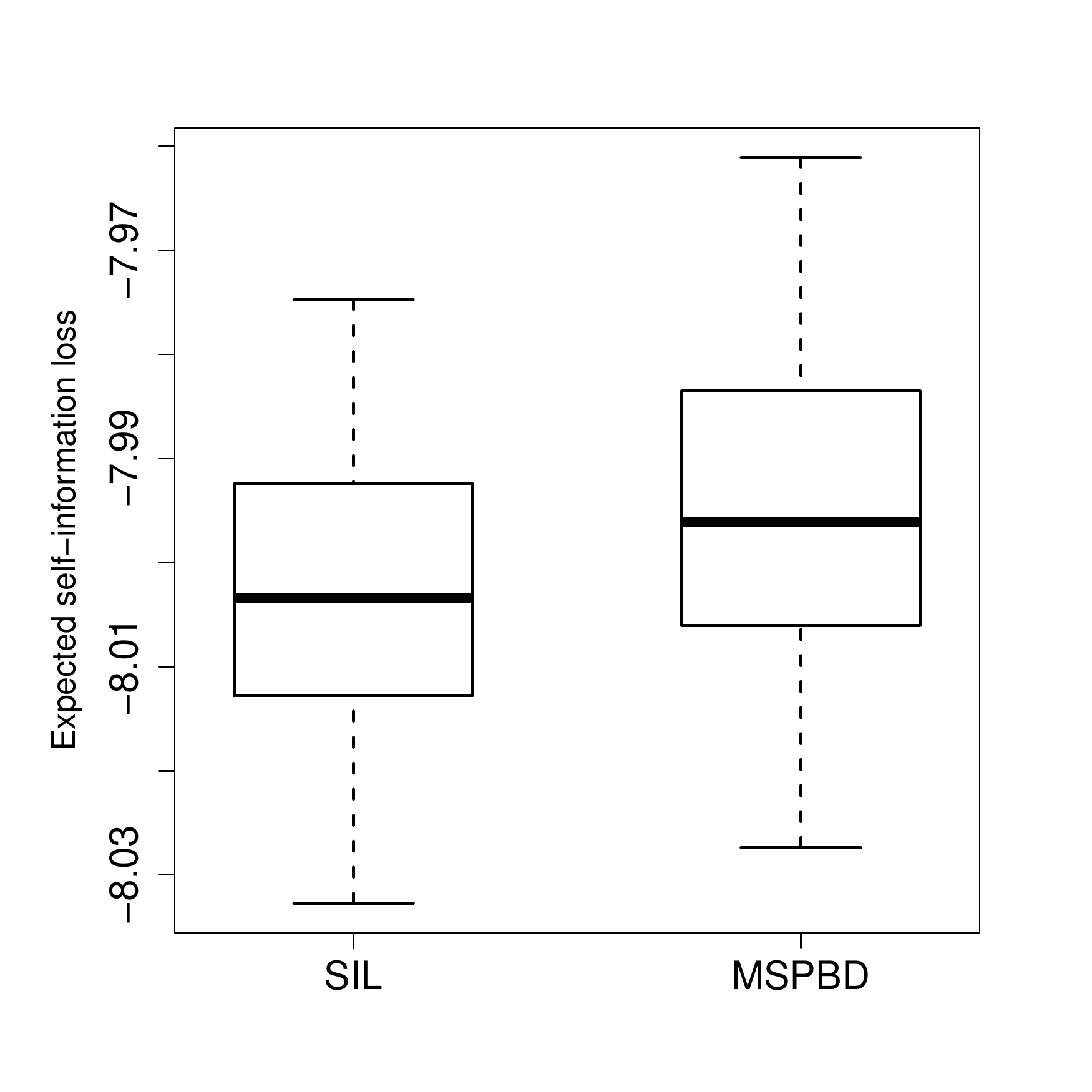}
\end{tabular} 
\end{center}
\caption{\label{fig:poissonresults}Log-linear regression example: boxplots of 20 Monte Carlo approximations of the negative expected KL divergence~\eqref{eq:NSIG} using $B=20,000$ simulations for the SIL-optimal and minimally supported Pseudo-Bayesian $D$-optimal designs (MSPBD) under uniform prior distributions~\eqref{eq:poisprior} for (a) $\alpha = 0.5$ and (b) $\alpha = 0.75$.}
\end{figure}

A quantitative comparison of the designs is given in Figures~\ref{fig:poissonresults}(a) for $\alpha=0.5$ and~\ref{fig:poissonresults}(b) for $\alpha=0.75$, which display 20 repeated Monte Carlo approximations of~\eqref{eq:NSIG}, each using $B=20,000$. The expected loss is lower for the less diffuse prior distribution ($\alpha = 0.5$). For both values of $\alpha$, the SIL-optimal and MSPBD-optimal designs perform very similarly, showing that the asymptotic approximation~\eqref{eq:dopt} is considerably more effective for this problem than for the binary response example, even though the experiment size $n$ is smaller. This finding is unsurprising as the normal distribution, from which~\eqref{eq:dopt} is derived, is a more effective approximation to  the Poisson distribution than it is to the Bernoulli distribution.

%\begin{figure}
%\begin{center}
%\includegraphics[scale=0.5]{figs/poisson_designs.pdf}
%\end{center}
%\caption{\label{fig:poisson_designs}Two-dimensional projections of the NSIG-optimal and minimally-supported Pseudo-Bayesian $D$-optimal designs.}
%\end{figure}

\subsection{Optimal designs for dimensional analysis}\label{sec:daresults}

\begin{figure}
\begin{floatrow}
\capbtabbox[\FBwidth]{%
 \begin{tabular}{ccccc}
\multicolumn{5}{c}{(a) SEL-optimal design} \\
Run & $x_1$ & $x_2$ & $x_3$ & $-\log\Pi_1$\\
\hline
1 & 0.070 & 0.079 & 0.095 & -1.753\\
2 & 0.070 & 0.087 & 0.102 & -1.843\\
3 & 0.070 & 0.076 & 0.116 & -1.769\\
4 & 0.076 & 0.089 & 0.073 & -1.593\\[7ex]
\multicolumn{5}{c}{(b) $V$-optimal design} \\
Run & $x_1$ & $x_2$ & $x_3$ & $-\log\Pi_1$\\
\hline
1 & 0.070 & 0.090 & 0.120 & -1.903 \\
2 & 0.070 & 0.090 & 0.120 & -1.903 \\
3 & 0.120 & 0.030 & 0.070 & 0.351 \\
4 & 0.120 & 0.030 & 0.070 & 0.351\\
\end{tabular}
}{%
  \caption{\label{tab:helidesigns}Helicopter experiment: (a) SEL-optimal and (b) $V$-optimal designs.}%
}
\ffigbox{%
 \includegraphics[scale=.4]{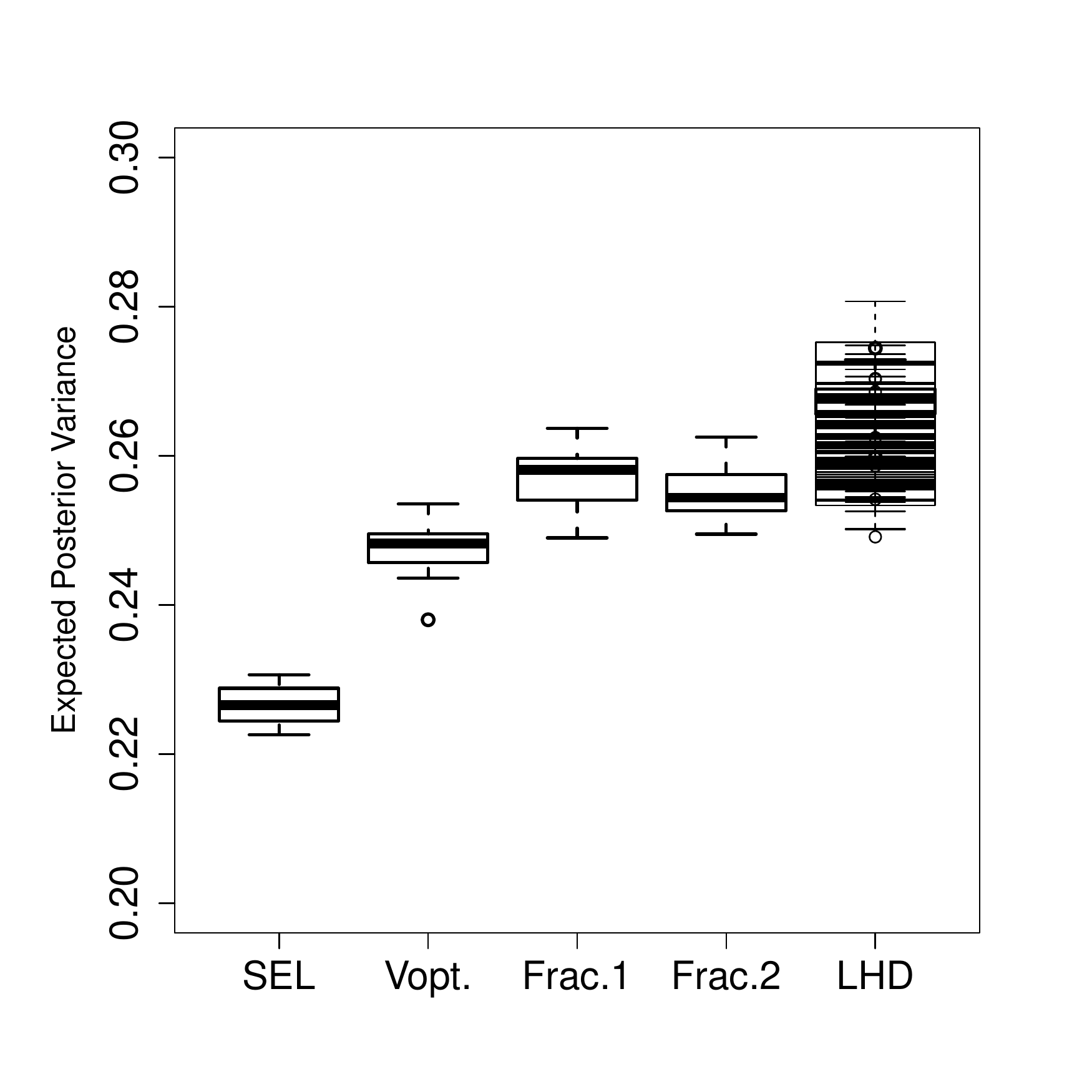}
}{%
\caption{\label{fig:heliresults}Helicopter experiment: boxplots of 20 Monte Carlo approximations of the average expected posterior variance~\eqref{eq:SEL} from $B=20,000$ simulations.}%
}
\end{floatrow}
\end{figure}

For the paper helicopter experiment, we find a SEL-optimal design, minimising a Monte Carlo approximation to~\eqref{eq:SEL}, for the Gamma regression model with linear predictor~\eqref{eq:helilinpred} and prior distribution~\eqref{eq:daprior}. The integral with respect to $\bx$ in~\eqref{eq:SEL} is approximated by summation across a $4^3$ grid of values constructed from $x_1,x_3\in\{0.07, 0.087, 0.103, 0.12\}$ (rotor and tail length) and $x_2\in\{0.03,0.05,0.07,0.09\}$ (rotor width). The design is given in Table~\ref{tab:helidesigns}(a) in terms of both the original three variables and the base quantity, $\Pi_1 \in[-1.903,0.351]$. Clearly, any design that results in the same values of $\Pi_1$ will have the same expected posterior variance. A goal of equally weighted prediction over $\mathcal{X}$ leads to greater weight being given to smaller values of $\Pi_1$, leading to the design including only points with smaller values of the base quantity.

We compare the SEL-optimal design to four competitors:
\begin{itemize}
\item A $V$-optimal design: the Fisher information matrix for $\bbeta$ for Gamma regression with the log link is equivalent to that for a linear model. Hence, classical optimal designs for the linear model (\citealp{ADT2007}, ch.~10) can be employed with this example. A $V$-optimal design, that minimizes the average prediction variance, has equally replicated design points  with $-\log\Pi_1 = -1.904$ and $-\log\Pi_1 = 0.351$. One such design is given in Table~\ref{tab:helidesigns}(b).   
\item The two regular $n=2^{3-1}$ fractional factorial designs with defining relation $I=x_1x_2x_3$.
\item A maximin Latin hypercube (LH) design (\citealp{SWN2003}, ch.~5) with $n=4$ points: twenty such designs were generated algorithmically using different starting designs. Each will be an approximation to the (globally) optimal maximin LH design.
\end{itemize}

A quantitative comparison of these designs is given in Figure~\ref{fig:heliresults} which displays boxplots from 20 Monte Carlo approximations to~\eqref{eq:SEL} for the SEL-optimal, $V$-optimal, fractional factorial and LH designs. The SEL-optimal design has a performance advantage over all the other designs, having average expected posterior variance around 8\% smaller than the $V$-optimal design, 11-12\% smaller than the fractional factorial designs, and 11-17\% smaller than the LH designs.

%\begin{figure}[htbp]
%\begin{center}
%\begin{tabular}{cc}
%(a) & (b) \\[-2ex]
%\includegraphics[scale=.4]{figs/heli_designs_pairs} & \includegraphics[scale=.4]{figs/heli_boxplot}
%\end{tabular}
%\end{center}
%\caption{\label{fig:heliresults}Competing designs for the helicopter experiment: (a) Two dimensional projections of the design points; (b) Boxplots of 20 Monte Carlo approximations of the expected posterior predictive variance from $B=20,000$ simulations.}
%\end{figure}

\section{Discussion}
\label{sec:disc}

Optimal Bayesian design is challenging for high-dimensional problems with multi-variable models and/or many design points and there are few literature examples of such designs being used in practice. Reasons for this include the lack of scaleable algorithms for design selection, the complexity of available software for Bayesian design, an occasional unwillingness to ``bias'' designs through the use of prior information (that is, to focus design performance on a certain set of parameter scenarios or models) and, in many application areas, a lack of appreciation that design of experiments can go beyond standard factorial designs. However, Bayesian design is a powerful tool for a variety of experiments. Here, we have focussed on using new methodology to find designs for generalised linear models where some prior information is \textit{necessary} to design informative experiments. We have also demonstrated for the first time how algorithmic Bayesian design can combine empirical and physical modelling principles via generalised linear models and dimensional analysis.

Methodology such as ACE removes some of the barriers to the implementation of Bayesian design, both by widening the scope of models and experiments that can be addressed, and by facilitating the provision of greater evidence for the effectiveness of the methods through rigorous scientific studies. Although the methodology is still computationally challenging for larger examples, an increase in statistical efficiency that allows even slightly smaller experiments to be run can lead to considerable cost savings in expensive industrial experimentation. These savings will often more than offset the additional computational time and resource used to find the designs. 

Clearly, in practice it is not always sensible to choose a design based solely on a one-number summary of design performance, particularly if it is obtained from a generic loss function that may not capture the aims of the experiment. However, being able to find optimal, or near-optimal, designs under suitable loss functions enables a short-list of competing designs to be compared on other merits. The methodology demonstrated in this paper allows the experimenter to understand any trade-offs resulting from the incorporation of other practical considerations.

More details of the methodology demonstrated in this paper can be found in \citet{OW2015} and also in \citet{OWP2015} who discussed optimal designs for uncertainty quantification of physical models, an application area of increasing importance. We demonstrated ACE using straightforward Monte Carlo approximations to the expected loss. However, the methodology can be applied with a variety of different approximations to the loss function, including asymptotic approximations to find pseudo-Bayesian designs. The ACE algorithm has been implemented in \texttt{R} package \texttt{acebayes} \citep{ace}.

\section*{Acknowledgements}
Woods was supported by Fellowship EP/J018317/1 from the UK Engineering and Physical Sciences Research Council. We are grateful to a reviewer for comments that improved the paper. 

\bibliographystyle{chicago}
\bibliography{mybib}
\end{document}